\documentclass[review,number,sort&compress]{elsarticle}
\usepackage{lineno}

\usepackage{amssymb}
\usepackage[ngerman,english]{babel}
\usepackage{lscape}
\usepackage{rotating}

\journal{Nuclear Physics A 	doi:10.1016/j.nima.2015.05.028 }

\begin{document}

\begin{frontmatter}



\title{ Characterization and performance of the ASIC (CITIROC) front-end  
of the ASTRI camera}


\author[label1]{D. Impiombato\corref{cor1}}
\ead{Domenico.Impiombato@iasf-palermo.inaf.it} 
\cortext[cor1]{Corresponding author. Tel.: +39 0916809468; fax: +39 0916882258}
\author[label1]{S. Giarrusso}
\ead{Giarrusso@iasf-palermo.inaf.it}
\author[label1]{T. Mineo}
\ead{Mineo@iasf-palermo.inaf.it}
\author[label1]{O. Catalano}
\ead{Catalano@iasf-palermo.inaf.it}
\author[label1]{C. Gargano}
\author[label1]{G. La Rosa}
\author[label1]{F. Russo}
\author[label1]{G. Sottile}
\author[label2]{S. Billotta}
\author[label2]{G. Bonanno}
\author[label2]{S. Garozzo}
\author[label2]{A. Grillo}
\author[label2]{D. Marano}
\author[label2]{G. Romeo}

 \address[label1]{INAF, Istituto di Astrofisica Spaziale e Fisica cosmica di Palermo, via U. La Malfa 153, I-90146 Palermo, Italy}
 \address[label2]{INAF, Osservatorio Astrofisico di Catania, via S. Sofia 78, I-95123 Catania, Italy}


\begin{abstract}
The Cherenkov Imaging Telescope Integrated Read Out Chip,
CITIROC, is a chip adopted as the front-end of the camera 
at the focal plane of the imaging Cherenkov ASTRI dual-mirror small size 
telescope (ASTRI SST-2M) prototype. 
This paper presents the results of the measurements performed to 
characterize CITIROC tailored for
the ASTRI SST-2M focal plane requirements.
In particular, we investigated the trigger linearity and efficiency,
 as a function of the pulse amplitude.
Moreover, we tested its response 
by performing a set of measurements using a silicon photomultiplier (SiPM)
in dark conditions and under light pulse illumination.
The CITIROC output signal is found to
vary linearly as a function of the input pulse amplitude. 
Our results show that it is suitable for the ASTRI SST-2M camera.

\end{abstract}

\begin{keyword}

 Front-end: ASIC for SiPM, ASTRI, CITIROC

\end{keyword}

\end{frontmatter}

\section{Introduction}
The global effort in designing and constructing
 the Cherenkov Telescope Array (CTA), which will have a 
ten times higher sensitivity than currently operating Cherenkov 
telescopes\cite{acharya13,actis}, is justified by 
the need of increasing the sensitivity in the
very-high energy (VHE) sky observations. 
Among CTA scientific goals, some pertain to the VHE regime, 
such as the understanding of the origin of cosmic rays, 
the nature of particle acceleration in shock regions
observed in supernova remnants and in black hole jets, 
and to unveil the enigmatic nature of dark matter. 

The Italian contribution to the CTA Project is mainly represented by 
the ASTRI (Astrofisica con Specchi a Tecnologia Replicante Italiana)
 program \cite{lapalombara13},
a flagship project currently financed by the Italian Ministry 
of Education, University and Research (MIUR) and led by 
the Italian National Institute for Astrophysics (INAF).
Its primary target is to develop an end-to-end 
prototype of the small-size class of telescopes (SST) 
devoted to the study of the highest gamma-ray energy 
range (from a few TeV up to 100 TeV and beyond).
The prototype, named ASTRI SST-2M, is characterized
by innovative technological solutions adopted for the first time
in the design of Cherenkov telescopes: the
optical system is arranged in a dual-mirror (2M) configuration
\cite{canestrari13,bonnoli13}, and the camera at 
the focal plane is composed of a
matrix of multi-pixel silicon photo-multipliers 
\cite{catalano13,catalano14,billotta14,bonanno13,bonanno14}.
\\
The telescope is based on a dual-mirror Schwarzschild-Couder 
design that allows for a compact optical configuration
with the ratio of 
the lens's focal length to the diameter of the entrance pupil
equal to 0.5.
The focal length is 2.15 m and the full field of view (FoV) is 9.6$^{\circ}$.
\\  
Using SiPMs instead of the traditional photo-multiplier tubes (PMTs)
offers advantages in terms of an excellent single
photon resolution, high photon detection efficiency (PDE), low
bias voltage (of the order of 70-73V), no damage due to
ambient light.  
The drawbacks in the version of the sensors 
used for ASTRI-SST 2M are: high dark rate ($>$ 1MHz)
 optical cross talk($>$ 20\%) and a gain which is strongly dependent on
temperature. Such drawbacks, however, do not prevent SiPMs from being
used as detectors at the focal plane of a Cherenkov telescope.
In fact, the dark count rate (DCR)($\sim$ 1MHz) is well below that of the night
sky background (NSB)($\sim$ 40MHz) and the gain 
can be kept stable with adequate feedback control of temperature 
 and over-voltage settings while optical cross talk is lower
than the intrinsic statistic fluctuation of the number of 
Cherenkov photons in air showers.

The very short (a few tens of ns) duration of the Cherenkov light flashes
associated with showers, requires a front-end electronics
(FEE) able to provide auto-trigger capability 
and fast camera pixel  read out.
The earlier FEE proposed for ASTRI was based on the extended analogue
silicon photo-multiplier integrated read out chip 
(EASIROC) \cite{marano13,callier11}, a commercial
application-specific integrated circuit (ASIC) for SiPM  read out.
Its characterization, performed with detailed 
measurements \cite{impiombato12,impiombato13},
proved that EASIROC fulfills the ASTRI requirements for 
the trigger time walk (5.5 ns), jitter (below 0.3 ns),  
electronic noise levels and electronic cross talk between 
channels.
Moreover, EASIROC is capable of providing auto-triggering as required by 
the very short duration of the air shower events. 
When a pixel detects a signal above the set threshold, 
a trigger is generated and sent to a Field Programmable Gate Array (FPGA)
that produces and sends back a hold 
signal (HOLD-$B$) to all 32 channels to stop the acquisition
and to start the output  read out.
At the hold time, directly set by the user using
a track-and-hold cell, the chip saves the amplitudes of the preamplified 
and the shaped signal.
Unfortunately, the ASIC shows in the high-gain (HG) 
electronics chain a drift  of 20-30ns in the peaking time of the shaped
signal that depends on the amount of injected charge.
In addition to that, the signal in a shower has an intrinsic duration
of the order of 10-30ns and the involved pixels are fired 
at different times.
When the camera trigger is set, all pixel signals are sampled at the same
time and then the two effects will produce a degradation of energy resolution
of the event, because 
the signals are read at different points of the shaping function.

To solve this problem, as explained later on, a peak detector circuit 
has been implemented in a new version of the EASIROC ASIC, named CITIROC
(Cherenkov Imaging Telescope Integrated Read Out Chip),
leaving the still suitable parameters unchanged.

In this paper, we present a set of measurements to
characterize the new functionalities introduced 
in CITIROC \cite{fleury14}. 
We evaluated the optical cross talk 
and the gain variation as a function of SiPM operating voltage 
and temperature.
In particular, we describe the ASTRI focal plane and the CITIROC device
in section 2 and 3, respectively.
The laboratory setup is presented in section 4 and the results of the 
CITIROC characterization are given in section 5.
The SiPM performance using CITIROC is presented in section 6 and 
our conclusions are discussed in section 7.

\section{The ASTRI SST-2M camera}
\label{camera}
The camera at the ASTRI SST-2M  focal plane is based on monolithic 
Hamamatsu SiPMs~S11828-3344m\footnote{http://www.hamamatsu.com/sp/hpe/HamamatsuNews/HEN111.pdf}
with 4$\times$4 squared physical pixels, 3$\times$3mm$^2$ each and made up
of 3600 elementary diodes of 50 $\mu$m pitch, yielding a filling factor 
of 62\%.
In order to match the angular resolution of the optical system, 
the physical pixels are grouped in 2$\times$2 logical 
pixels (6.2$\times$6.2mm$^2$) with a sky-projected angular size of 
0.17$^{\circ}$ that includes 80\% of the optical 
point spread function (PSF).
A complete characterization of the ASTRI SST-2M SiPMs 
using physical pixels is reported in \cite{marano14}.
Considering the size of the optical area of ASTRI SST-2M,
energy range from a few TeV up to 100 TeV and beyond,  
the requirement for the maximum number of photoelectrons (pe)
detected in one pixel is 1000 with a goal of 2000.

\begin{figure*}[ht!!]
\centering
\includegraphics[angle=0, width=11cm]{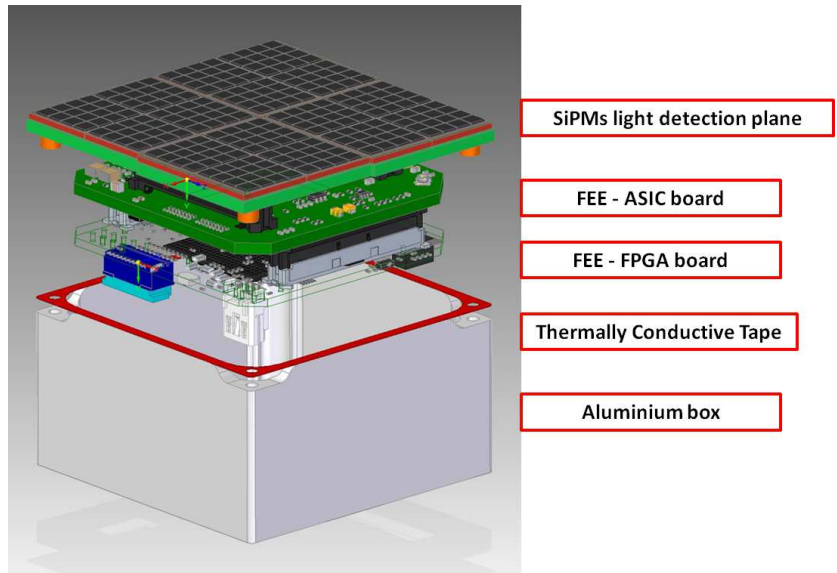}
\caption{A schematic view of the PDM mechanical assembly.}
\label{fig1}
\end{figure*}

\begin{figure*}[ht!!]
\centering
\includegraphics[angle=0, width=11cm]{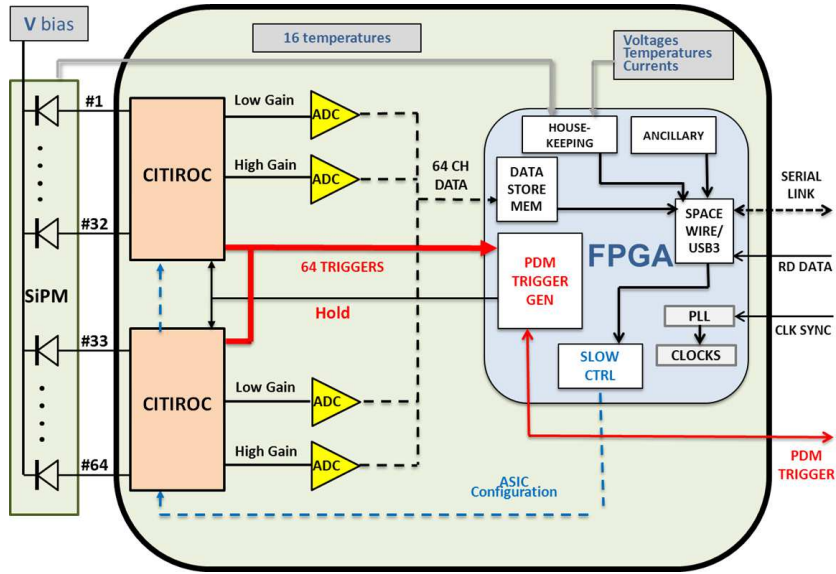}
\caption{A schematic view of the connections to the ASICs.}
\label{fig2}
\end{figure*}

For modularity and fast  read out of the focal plane, 
the detector units are organized in 
an array of 37 photon detection modules (PDM), 
each with 8$\times$8 logical pixels (see Fig.~\ref{fig1}). 
Two CITIROC ASICs are then necessary to read a single PDM,
since each CITIROC chip contains 32 channels (see Fig.~\ref{fig2}).

The telescope trigger is activated when $n$ contiguous pixels
within a PDM  present a signal above the discriminator threshold.
 
The number $n$ of contiguous pixels and the trigger threshold are chosen
in order to have a maximum rate of $\sim$600 Hz  from the whole camera, 
ensuring a dead time $<$3\%, according to the CTA requirements.
From simulations, we evaluated that 5 contiguous pixels with
a discriminator threshold above 4 pe in each pixel
gives an average rate lower than requirement.

\section{The CITIROC chip}
\begin{figure*}[ht!!]
\centering
\includegraphics[angle=0, width=10cm]{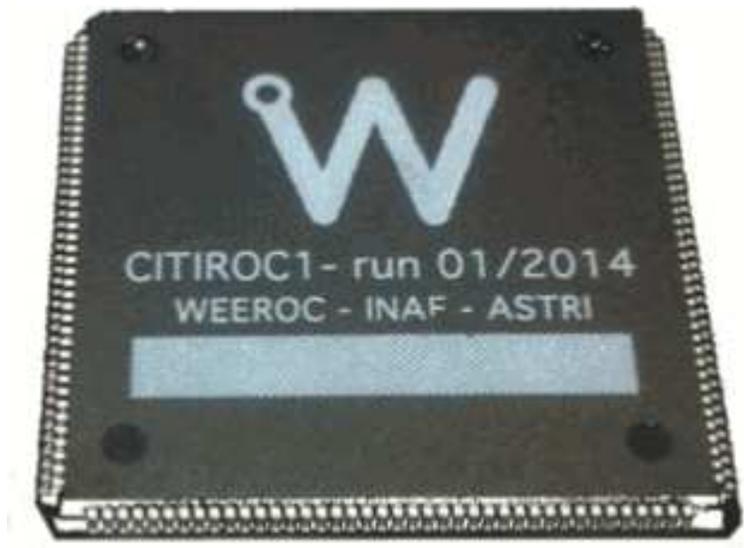}
\caption{CITIROC chip 4.9$\times$4.6mm$^2$ (courtesy of WEEROC).}
\label{fig3}
\end{figure*}

\begin{figure*}[ht!!]
\centering
\includegraphics[angle=0, width=16cm]{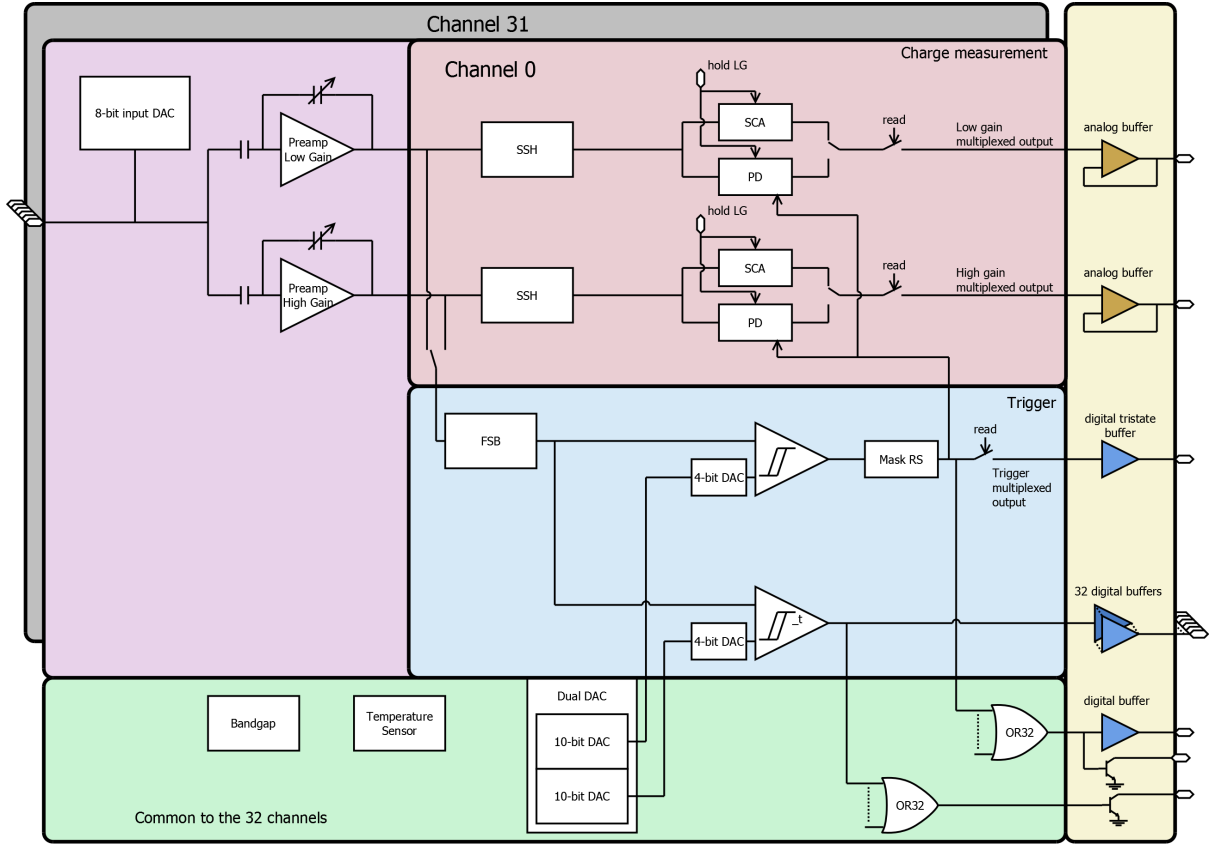}
\caption{Architecture of the front-end CITIROC (courtesy of WEEROC).}
\label{fig4}
\end{figure*}

CITIROC, designed by WEEROC\footnote{http://www.weeroc.com}, is based 
on the EASIROC architecture with the addition 
of custom functions as required by the ASTRI team (see Fig.~\ref{fig3}).
It is a 32 channel fully analog front-end  ASIC designed to  read out SiPMs 
for single photon detection. 
Its analog core and characteristics are shown in Fig.~\ref{fig4} 
and listed in Table~\ref{table1}, respectively.
The  processing of the analog signal takes place in the front-end channels of the device, 
while the  read out is handled at the internal back-end of the ASIC.

Two separate electronics chains allow for
high- and low-gain (HG and LG) simultaneous processing of the analog signal.

Each of the two chains is composed (see Fig.~\ref{fig4}) of an adjustable preamplifier 
followed by a tunable shaper (SSH: slow shaper), a track-and-hold circuit (SCA: switched capacitor array) and an 
active peak detector (PD: peak detector) to capture and hold the 
maximum value of signal.
Fine-tuning of each pixel gain is obtained adjusting
the voltage applied to the SiPM through an 8-bit 
digital-to-analog converter (DAC) ranging from 0 to 4.5 V.

A third chain implements 
the trigger channel generation using a fast shaper 
(FSB: bipolar fast shaper) 
with fixed shaping 
time of 15 ns, followed by two discriminators.
A 10-bit DAC common to all 32 channels, sets the programmable 
threshold to the discriminators (see Fig.~\ref{fig4}).
The first of two discriminators, that can also be masked,
gives a single output for all channels, while the second 
one provides a different output for each channel.
The topological trigger (see Sect.~\ref{camera}),
implemented in the ASTRI SST-2M camera, can be obtained from the second.
All CITIROC main parameters can be programmed by downloading a 
configuration bit-string through a slow-control serial line.
The outputs of all the channels can be  read out by multiplexing, 
in parallel, the analog buffers of the HG and LG chains.

An evaluation board has been designed by WEEROC
to test the functional characteristics and performance of the ASIC. 
It allows easy access to the CITIROC outputs and provides many test 
points to the FPGA\footnote{Altera Corporation - Cyclone FPGA Model EP3C16Q240C8N} 
dedicated lines. 
 It is equipped with two external analog-to-digital converter (ADC) 
to allow the reading of the analog processed data by the ASIC in 
digital form.  

 A Lab-VIEW\footnote{http://www.ni.com/labview/i/} software procedure, 
 developed by the LAL 
(Laboratoire de l'Acc\'el\'erateur Lin\'eaire) 
Tests group\footnote{ http://www.lal.in2p3.fr/}, has been provided, 
together with the evaluation board, to command 
 the CITIROC chip and receive the outputs via universal serial bus 
 (USB) connection. 

\setcounter{table}{0}
\begin{table*}[ht]
\centering

\caption{Main characteristics of the CITIROC chip.}
\label{table1}
\begin{tabular}{|l|l|}
\hline
Technology:     & Austria-Micro-Systems (AMS) SiGe 0.35 $\mu$m\\
\hline
Dimensions :    &    16.5 mm$^2$ (4.1$\times$4.1mm)  \\
 \hline
Power Supply : & 4.5 V/0 V \\
 \hline
Consumption:    & 2 mW per channel\\
 \hline
                & 95 mW in data transmission mode (all outputs on)\\
 \hline
Inputs:  	& 32 voltage inputs with independent SiPM HV adjustments\\
\hline
Outputs: 	& 32 trigger outputs \\
 \hline
            & 1 multiplexed charge output\\
 \hline
            & 1 ASIC trigger output (Trigger OR)\\
 \hline

Internal Programmable Features:  &32 HV adjustments (one for each channel) (32x8bits)\\
 \hline
                                 &Trigger Threshold Adjustment (common for all channels) (10bits)\\
\hline
                                 &Gain tuning (channel by channel)       \\
\hline
                                 &32 Trigger Masks                     \\
\hline
                                 &Channel by channel output enable     \\
\hline

\hline
Package :       & Naked (PEBS) TQFP160  \\
\hline
 \end{tabular}
 \end{table*}

\clearpage
\newpage

\section{Laboratory set-up}
We performed two different sets of measurements:
the first had the aim to characterize the new circuits of CITIROC and
the second was performed to evaluate the optical cross talk
and the gain variation for the SiPM configuration with the logical pixels
adopted for the ASTRI SST-2M camera. 
The gain of the LG chain preamplifier is set to 5 
to produce a monotonic response up to 2000 pe and 
the gain of the HG chain preamplifier is set to 150
allowing an almost linear working range up to $\sim$100 pe (see Sect.~\ref{Peak_Detector}).
The shaping time constant of the signal is fixed to 37.5 ns, 
producing a nominal peaking time of $\sim$50 ns.

For the CITIROC characterization we used an arbitrary pulse-function generator 
to create a precise input charge.
The shape of the signal has been tailored to the actual shape of the SiPM
waveform. It is characterized by a very fast rise time
(a few hundreds of ps) followed by an exponential decay ($\sim$175 ns).
 The amplitude of this signal is 103$\pm$4 $\mu$V  for an input charge 
of 0.12 pC, equivalent to 1 pe for a SiPM operating at a gain of $7.5\times10^{5}$.

\begin{figure*}[h!!]
\centering
\includegraphics[angle=0, width=15cm,height=12cm]{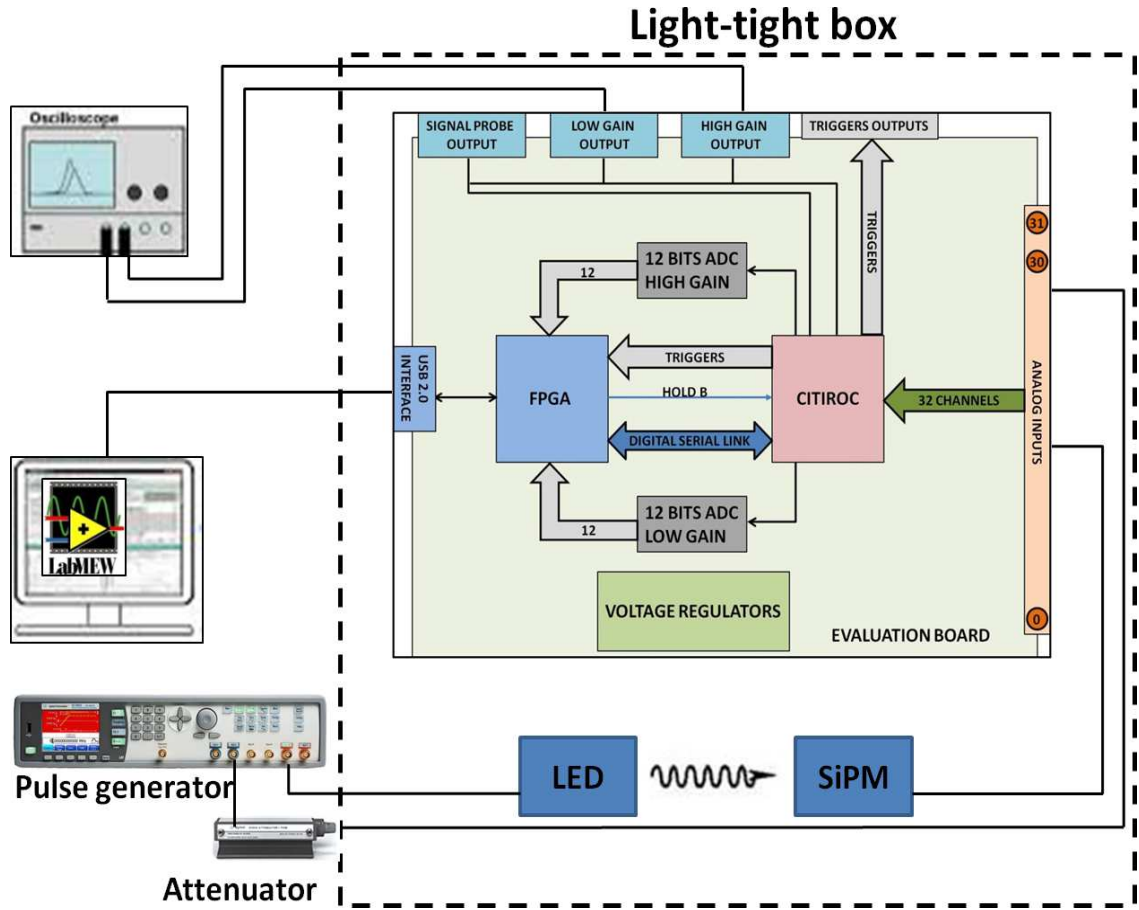}
\caption{Schematic view of the set-up used for the SiPM$+$CITIROC characterization.}
\label{fig5}
\end{figure*}

The second set of measurements uses the setup shown 
in Fig.~\ref{fig5}:
the SiPM and the front-end electronics are located 
in a small metal box which is placed inside a
controlled temperature chamber. 
The gain is finely tuned changing the operating voltage
through a DAC in steps of a few millivolt
in a range 71.60-72V. 
\\
For these measurements we used a light pulse,
emitted by a blue light emitting diode (B-LED) facing the 
SiPM pixels driven by the pulse generator.

All measurements presented in this paper were performed using only one 
of the 32 channels available in the CITIROC.

\newpage

\section{CITIROC characterization}
In this set of measurements, we compared the CITIROC energy resolution 
with that of EASIROC and investigated the response linearity.
Moreover, we measured the 
trigger efficiency as a function of the input charge
to complete the characterization published in \cite{impiombato13}.

\subsection{Performance of the Peak detector}
\label{Peak_Detector}
\subsubsection{Energy resolution}

\begin{figure*}[h!!]
\centering
\includegraphics[angle=0, width=10cm]{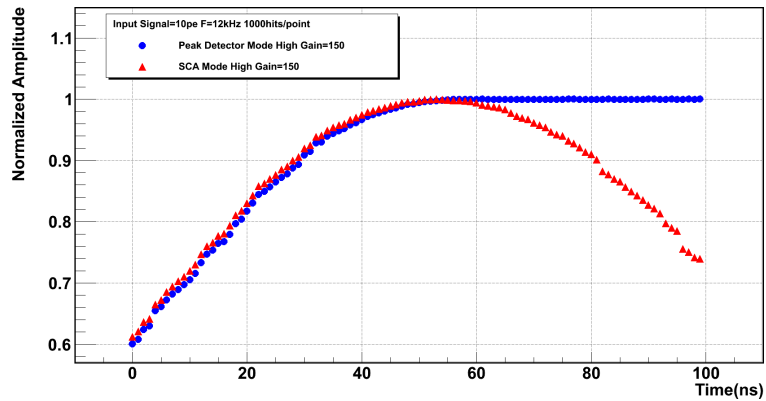}
\caption{Normalized amplitude of HG shaping function versus 
time in ns. The red triangles are relative to the SCA mode, and
the blue dots show the peak detector mode.}
\label{fig6}
\end{figure*}

\begin{figure*}[ht!!]
\centering
\includegraphics[angle=0, width=10cm]{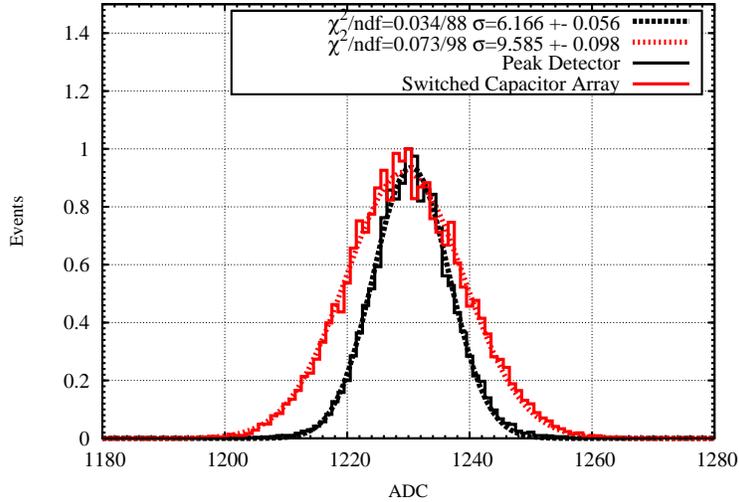}
\caption{Normalized histograms for injected charge of 10 pe obtained with
the peak detector (continuous black line) and the switched capacitor array
(continuous red line). The dashed lines represent the fitting functions.}
\label{fig7}
\end{figure*}

To reduce the degradation of the energy resolution 
due to incoherent sampling of the signal in EASIROC 
switched capacitor array (SCA) mode,
we introduced also an mode based on the peak detector circuit.
As shown in figure~\ref{fig6}, this mode,  at
difference from SCA,
keeps the maximum value once it has been reached.
To test its performance, 
we compared the 10 pe pulse-height distributions 
obtained with CITIROC with those given by EASIROC.
The distributions are generated sampling the HG shaped signal from 55 ns
to 70 ns in steps of 2 ns with 5000 events per run:
results are shown in Fig.~\ref{fig7}.
Fitting the two curves with Gaussians,
we find that the sigma of the peak detector distribution (6.17$\pm$0.06)
is significantly lower than that of the SCA (9.6$\pm$0.1) and 
is compatible with the intrinsic energy resolution (6.20$\pm$0.06) 
of the SiPM.  

\newpage
\subsubsection{Linearity}
\begin{figure*}[ht!!]
\centering
\includegraphics[angle=0, width=10cm]{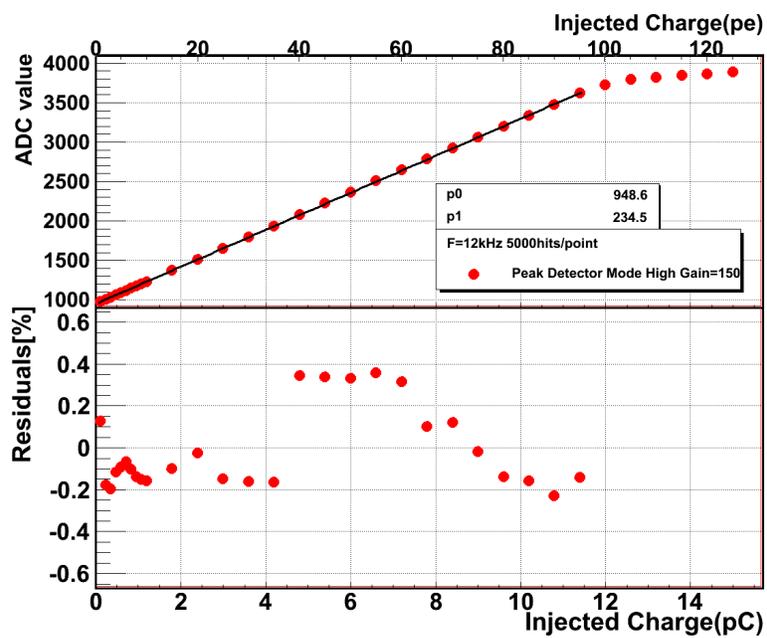}
\caption{ADC values versus injected charges for the HG chain.
The black line is the linear fit up to 95 pe where the ADC value saturates.
The p0 and p1 are the coefficients of the best fit.
The bottom panel shows the residuals
 respect to the line.}
\label{fig8}
\end{figure*}

\begin{figure*}[h!!]
\centering
\includegraphics[angle=0, width=10cm]{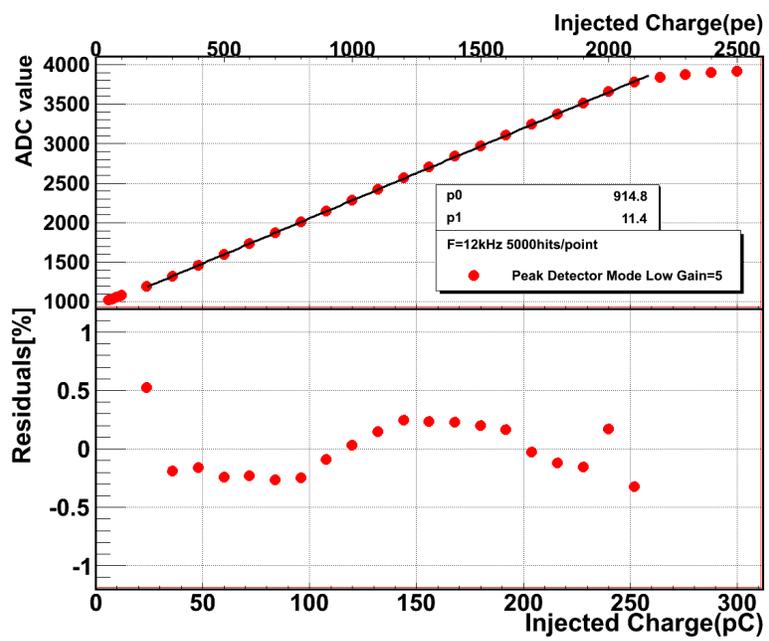}
\caption{ADC values versus injected charges for the LG chain.
The black line is the linear fit in the range 200-2100pe.
The p0 and p1 are the coefficients of the best fit.
The bottom panel shows the residuals
 respect to the line.}
\label{fig9}
\end{figure*}

We tested the linearity of the response in the HG and LG chains as a function 
of the input charge by varying it from 0.12 to 15 pC (1-125pe) for the HG and 
from 6 to 300 pC (50-2500pe) for the LG. 
For each point, we collected the distributions of the sampled signal 
over 5000 tests.
The ADC values and the errors related to each charge are computed from 
the average and the sigma of the distributions. 
Results are shown in Figures~\ref{fig8} and ~\ref{fig9}. 
We note that the ADC curves are linear up to 
11.4 pC (95 pe) in the HG chain and from 24 pC up to 252 pC (200-2100pe) 
in the LG chain, with residuals lower than 1\%.

\newpage

\clearpage
\newpage
\subsection{Trigger linearity and efficiency}
\begin{figure*}[h!!]
\centering
\includegraphics[angle=0, width=10cm]{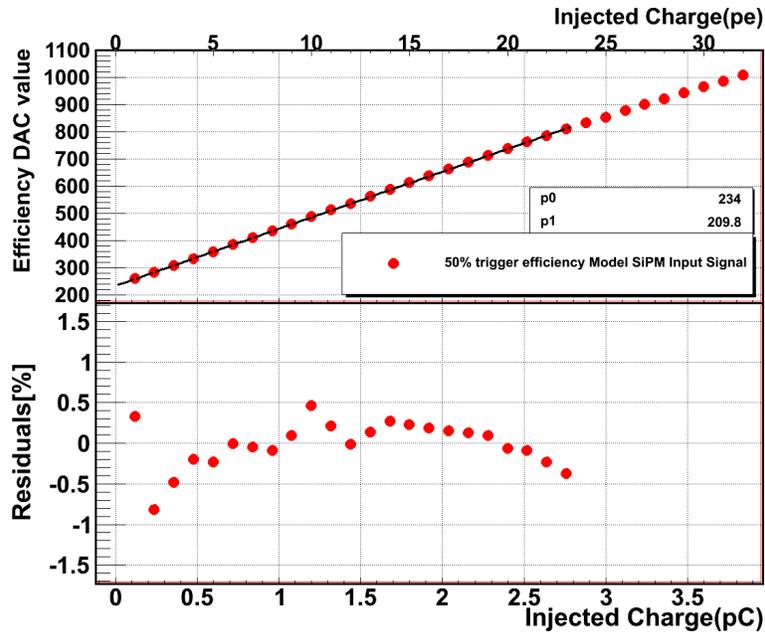}
\caption{50$\%$ trigger efficiency threshold versus injected 
charge from 0.12 to 3.84 pC. The p0 and p1 are the coefficients of 
the best fit. The bottom panel shows the residuals
 respect to the line. }
\label{fig10}
\end{figure*}

To check the linearity 
of the ASIC discriminator, we studied the curves of the trigger efficiency 
as a function of threshold. They are obtained by 
varying the input charge and the threshold level  while the 
other parameters, such as the preamplifier gain and shaping time, 
are kept constant.
Figure~\ref{fig10} shows the evolution 
of 50$\%$ of trigger efficiency as a function of the injected charge
in the range 0.12-3.84pC (1-32pe).
The trigger efficiency is linear, with a discrepancy 
lower than 1\% up to 2.64 pC (22 pe) and it saturates at 32 pe.

\section{Characterization of the SiPM configuration with logical pixel}
The SiPM physical pixel (3$\times$3mm$^2$) 
characterization has been already done and the 
results have been published in \cite{marano14}.
The measurements presented in this paper refer to 
the logical pixels configuration (6.2$\times$6.2mm$^2$)
adopted for the ASTRI SST-2M camera.

\subsection{Optical cross talk}

The optical cross talk level was evaluated by measuring 
the dark count rate as a function of the discriminator threshold. 
Assuming a Poisson distribution, the probability to 
have two coincident events within a time
window of 15 ns is about 0.001\% for the laboratory measured rate 
of 1 MHz: all events with a number of pe higher than 1 are due to optical 
cross talk.
The dark count rate drops when integer multiples of 1 pe thresholds 
are reached: the total rate is obtained
with thresholds above the electronic noise but below 1 pe; 
conventionally it refers to 1/2 pe. 
The characteristic curve obtained varying the discriminator threshold
is known as ''staircase''.

\begin{figure*}[h!!]
\centering
\includegraphics[angle=0, width=10cm]{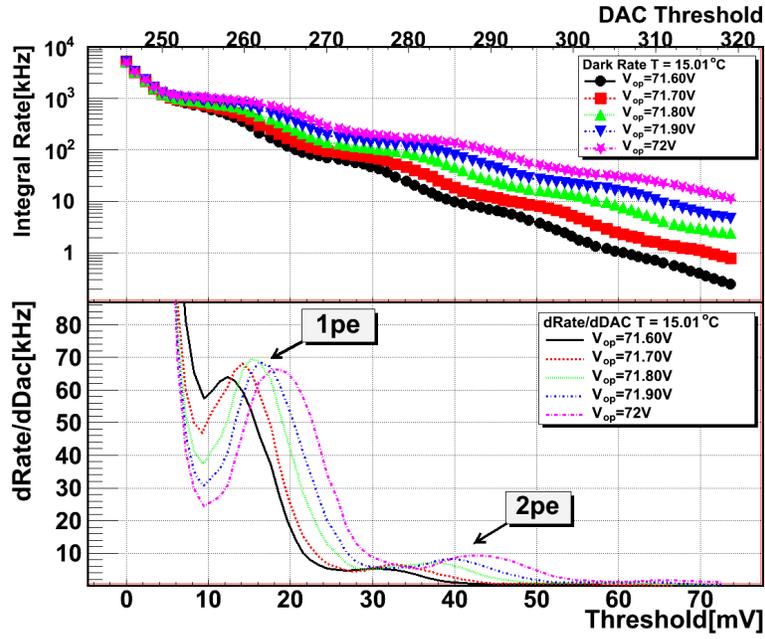}
\caption{{\it Top panel:} Dark rate as a function 
of the discriminator threshold at the 
fixed temperature of 15.01$^\circ$C and at different operating voltages 
values.
{\it Bottom panel:} First derivative of the dark rate vs 
the discriminator threshold, the arrows 
indicate the position where $\nu_{1pe}$ and $\nu_{2pe}$ are computed.}
\label{fig11}
\end{figure*}

In this set of measurements, we varied the operating voltage  
in the range 71.60-72.00V in steps of 100 mV,
keeping the temperature constant
at 15.0$^\circ$C$\pm$0.1$^\circ$C.
We accumulated the staircases for several operating voltages
as shown in the top panel of Figure~\ref{fig11}. 
The optical cross talk probability is measured with 
the ratio between the rate relative to 1 pe ($\nu_{1 pe}$) and
that corresponding to 2 pe ($\nu_{2 pe}$), 
$P_{c} = \nu_{2 pe}/\nu_{1 pe}$,
following the method applied in a first evaluation with 
the front-end EASIROC \cite{impiombato14}.  
The discriminator thresholds to evaluate $\nu_{1pe}$ and $\nu_{2pe}$
are obtained computing the first derivative of the staircases 
and evaluating the maxima of the curves 
by fitting the data with a spline function.

\newpage
\begin{figure*}[ht!!]
\centering
\includegraphics[angle=0, width=10cm]{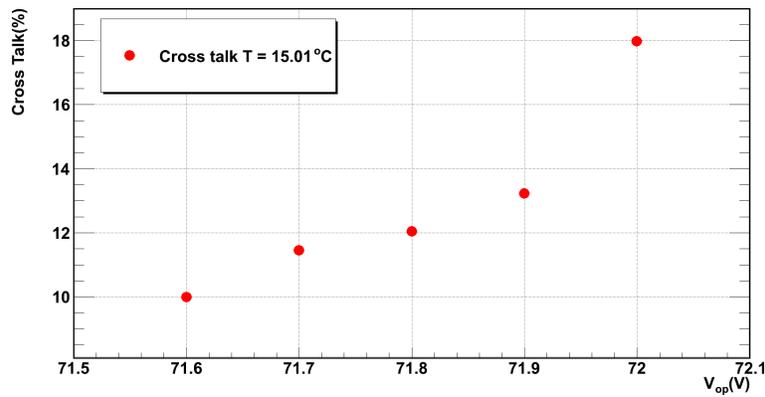}
\caption{Optical cross talk probability versus operating voltage at 15.01$^\circ$C.
The error of each point is less than 0.1 \%, so it is not visible 
in the figure}
\label{fig12}
\end{figure*}

As a result, figure~\ref{fig12} presents the optical cross talk values as a function 
of the operating voltage:
it increases from 10\% to 18\% in the investigated range.
These values are comparable to those 
measured in the SiPM characterization presented in \cite{marano14} 
for the physical pixels.

\subsection{Gain variation as function of temperature and voltage}
\begin{figure*}[h!!]
\centering
\includegraphics[angle=0, width=10cm]{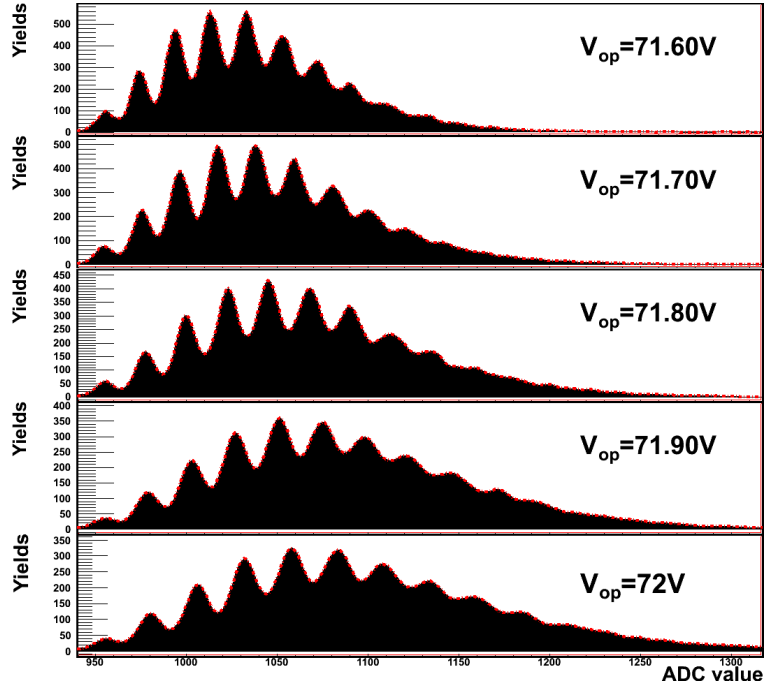}
\caption{Pulse height distributions for operating voltages in the 
range (71.60-72.00)V, computed at temperature T=15.01C$^\circ$C.}
\label{fig13}
\end{figure*}

To evaluate the gain, we used 
the pulse height distribution measured by pulsing a 
B-LED at a constant rate of 10 kHz.
The duration of the B-LED pulse was set in order to have
an average number of pe $\sim$4.
All results are then obtained for the HG chain.
Figure~\ref{fig13} shows the integrated charge spectra
at the investigated voltage values accumulated keeping the
temperature at 15.01$^\circ$C.
The first peak, referred to as "pedestal", includes events due to 
electronic noise and SiPM noise.
The gain $G$ in each histogram is determined  
from the average of the ADC distances between subsequent local maxima 
applying the following equation : 
\begin{equation}
\label{fm1}
 G\,=\,\frac{1}{N}\cdot\sum_{i=1}^{N} (ADC_{i+1}-ADC_{i})
\end{equation} 
\noindent 
where $G$ is given in ADC units, ADC$_{i}$ is the ADC value of the 
maximum of the i-th peak on the 
histogram while N is the number of peaks.
The error on $G$ is given by the standard deviation of  the measured
value with respect to the mean.

\begin{figure*}[h!!]
\centering
\includegraphics[angle=0, width=10cm]{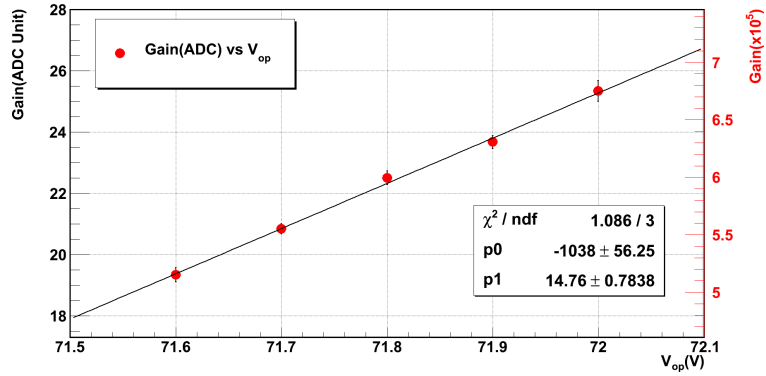}
\caption{Gain variation in ADC units as
 a function of the V$_{op}$ at 15.01$^\circ$C.
For clarity, the correspondent absolute value of the gain is reported 
on the right.}
 
\label{fig14}
\end{figure*}

We investigated the variation of the gain with voltage ($dG/dV$)
in the range 71.60-72.00V.
Using a linear fit for all the data we obtain the following 
relation (see figure~\ref{fig14}):
 
\begin{equation}
\label{fm2}
 \frac{dG_{ADC}}{dV}\,=\,14.76~~~~~~[ADC/V]
\end{equation}
\noindent 
where the gain is expressed in unit of ADC code.

\begin{figure*}[h!!]
\centering
\includegraphics[angle=0, width=10cm]{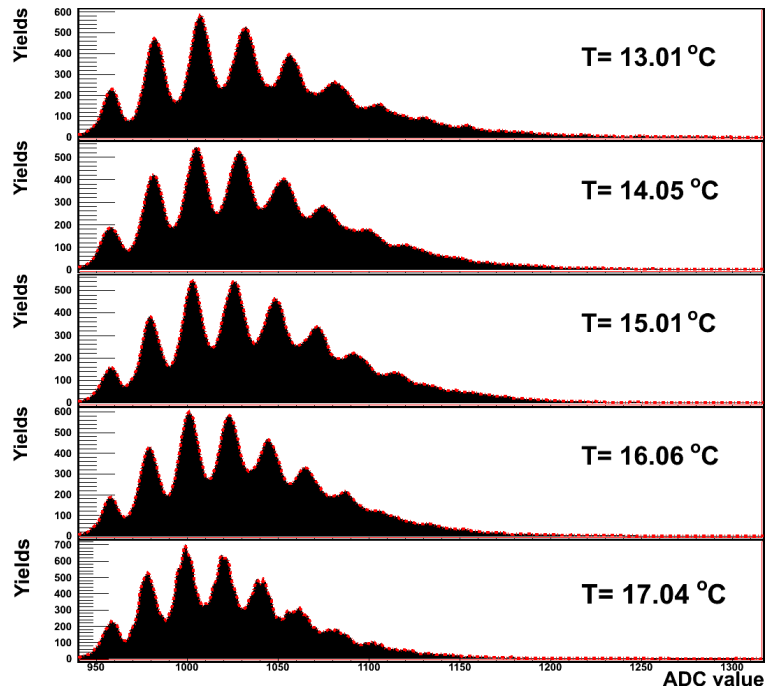}
\caption{Pulse height distributions for several temperatures 
at the operating voltage of 71.80 V.}
\label{fig15}
\end{figure*}

\newpage
Considering that the electronics does not introduce any variation
 with temperature, as we tested in laboratory, 
we also investigated the SiPM gain dependence on the temperature ($dG/dT$)
with the same method described above. 
Figure~\ref{fig15} shows the resulting pulse height distributions
obtained varying
the temperature from 13$^{\circ}$C to 17$^{\circ}$C for a fixed 
operating voltage of 71.80 V.
 
\begin{figure*}[ht!!]
\centering
\includegraphics[angle=0, width=10cm]{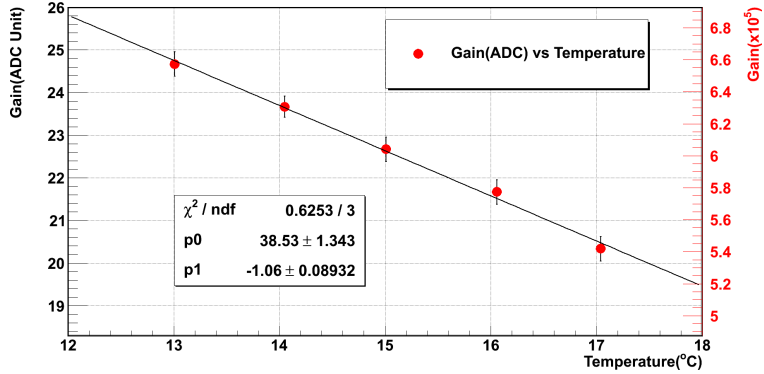}
\caption{Gain variation in ADC unit as
 a function of the temperature at 71.80 V.
 For clarity, the correspondent absolute value of the gain is reported 
on the right.}
\label{fig16}
\end{figure*}

The points are well inside a linear relationship as shown 
in figure~\ref{fig16};
the slope is given by the following equation: 
\begin{equation}
\label{fm3}
 \frac{dG_{ADC}}{dT}\,=\,-1.06~~~~~~[ADC/^{\circ}C]
\end{equation}
 \noindent 
where the gain is expressed in unit of ADC code.

\clearpage
 \newpage

\section{Conclusions}

We presented a set of measurements aimed at characterizing the peak detector
circuit implemented in CITIROC.
Moreover, we evaluated the level of optical cross talk and the rate of
the gain variations as a function of temperature and operating
voltage for the SiPMs Hamamatsu~S11828-3344m adopted for the 
ASTRI SST-2M camera.
Our results can be summarized in the following points:

\begin {itemize}

\item CITIROC overcomes the problems introduced by the SCA mode
and allows for coherent sampling of the pixels fired by the shower even
in case of large ($\sim$30 ns) image time gradient.
This results in improved energy resolution. 

\item The signal is linear with respect to the input charge up to 100 pe in HG
and up to 2000 pe in the LG.
The trigger dynamic range extends linearly up to 17 pe and saturates at 32 pe.
\item The SiPM Hamamatsu~S11828-3344m 
adopted for ASTRI SST-2M has an optical cross talk  level in the 
range 10-20\%. 
ASTRI SST-2M will be operated with SiPM gains
resulting in an optical cross talk lower than 15\% even if
this implies a loss in the linearity for low input charges.

The updated versions of the sensor foreseen for the ASTRI/CTA mini-array 
are expected to have a level of optical cross talk of 
a few percent.

\item We evaluated the function that models the variation of the gain 
with respect to voltage and temperature.
We find a good linearity in the investigated range: the gain 
increases by 14.76 ADC per volt, and decreases by 1.06 ADC for an increase
of one degree in temperature. 

\end {itemize}
We can conclude that the CITIROC ASIC
is fully suitable for ASTRI SST-2M
and it can also be adopted in any new version
of SiPM that could be proposed for ASTRI/CTA mini-array.

\section{Acknowledgements}

The work presented in this paper was supported in part by the ASTRI,
"Flagship Project" financed by the Italian Ministry of Education, University,
and Research (MIUR) and  led by the Italian National Institute 
for Astrophysics (INAF). 
We also acknowledge partial support from MIUR Bando PRIN 2009 and 
TeChe.it 2014 Special Grants.
We are deeply grateful to S. Callier, C. De La Taille, J. Fleury
and L. Raux within WEEROC at Orsay and to the
ASTRI collaborators for useful discussions and suggestions.
D.I., S.G., T.M. and O.C. thank M. Fiorini, E. Giro, M.C. Maccarone, S. Vercellone and
S. Wakley for their contributions as internal referees 
of the paper.







\bibliographystyle{model1c-num-names}
\bibliography{<your-bib-database>}

\begin{thebibliography}{99}



\bibitem{acharya13} Acharya, B. S. et al., ''Introducing the CTA concept'', Astroparticle Physics 43, 3 (2013).
\bibitem{actis} Actis, M. et al., ''Design concepts for the Cherenkov Telescope Array CTA: an advanced facility for groundbased
high-energy gamma-ray astronomy'', Experimental Astronomy 32, 193 (2011). arXiv:1008.3703.
\bibitem{lapalombara13}La Palombara, N. et al., for the ASTRI Collaboration, ''The INAF ASTRI Project in the framework of CTA'',
in: Procedings of the ICATPP-2013 Como, World Scientific, 2014 ArXiv:1405.4187.
\bibitem{canestrari13} Canestrari, R. et al., for the ASTRI Collaboration and the CTA Consortium, ''The ASTRI SST-2M prototype: structure and mirror'', in: Proceedings 33$^{rd}$ of the ICRC2013, 2-9 July 2013, Rio de Janeiro Brazil, (arXiv:1307.4851).
\bibitem{bonnoli13} Bonnoli, R. et al., for the ASTRI Collaboration and the CTA Consortium, ''Boosting the performance of the ASTRI SST-2M prototype: reflective and anti-reflective coatings'', in Proceedings 33$^{rd}$ of the ICRC2013, 2-9 July 2013, Rio de Janeiro Brazil, (arXiv:1307.5405).
\bibitem{catalano13} Catalano, O. et al., For the ASTRI collaboration, ''The ASTRI SST-2M prototype: camera and electronics'', in: Proceedings ICRC 2013, Rio De Janeiro, Brazil, July 2013.
\bibitem{catalano14} Catalano, O. et al., For the ASTRI Collaboration and the CTA Consortium, ''The camera of the ASTRI SST-2M prototype for the Cherenkov telescope array'', in: Proceedings SPIE 2014  Ground-Based and Airborne Instrumentation for Astronomy, vol. 91470D, Montreal, Canada, June 2014.
\bibitem{billotta14} Billotta, S. et al., for the ASTRI Collaboration and the CTA Consortium, ''SiPM detectors for the ASTRI project in the framework of the Cherenkov telescope array'', in: Proceedings SPIE 2014  High Energy  Optical  and Infrared Detectors for Astronomy VI, vol. 91541R, Montreal, Canada, July 2014.
\bibitem{bonanno13} Bonanno, G. et al., ''Characterization Measurements Methodology and Instrumental Set-up Optimization for New SiPM Detectors - Part I: 
Electrical Tests'', IEEE Sensors Journal 14, 10 (October), 2014, 3557.
\bibitem{bonanno14} Bonanno, G. et al., ''Characterization Measurements Methodology and Instrumental Set-up Optimization for New SiPM Detectors - Part II: 
Optical Tests'', IEEE Sensors Journal 14, 10 (October), 2014, 3567.
\bibitem{marano13} Marano, D. et al., ''PSPICE High-Level Model and Simulations of the EASIROC Analog Front-End'', International Journal of Modelling and Simulations 34 (4), 2014.
\bibitem{callier11} Callier, S. et al., ''EASIROC, an easy \& versatile  readout device for SiPM'', TIPP 2011 – Technology and Instrumentation in Particle Physics 2011.
\bibitem{impiombato12} Impiombato, D. et al., on behalf of the ASTRI Collaboration, ''Characterization of the front-end EASIROC for  readout of SiPM in the ASTRI camera'', in: Proceedings of the SciNeGHE 2012 Workshop, 20-22 June 2012, Lecce, Italy, (Nuclear Physics B (Proceedings Supplements) 239 (June 2013) 254).
\bibitem{impiombato13} Impiombato, D. et al., on behalf of the ASTRI Collaboration, ''Characterization of EASIROC as front-end for the  readout of the SiPM at the focal plane of the Cherenkov telescope ASTRI'' Nuclear Instruments and Methods in Physics Research Section A 729 (November), 2013, 484.
\bibitem{fleury14} Fleury, J. et al., ''Petiroc and Citiroc : Front-end ASICs for SiPM  readout and ToF applications'', Journal of Instrumentation 9, 2014, C01049.
\bibitem{marano14} Marano, D. et al., ''Electro-optical characterization of MPPC detectors for the ASTRI Cherenkov telescope camera'', Nuclear Instruments and Methods in Physics Research Section A 768, 2014, 32.
\bibitem{impiombato14} Impiombato, D. et al., on behalf of the ASTRI Collaboration, ''Evaluation of the optical cross talk level in the SiPMs adopted in ASTRI SST-2M Cherenkov Camera using EASIROC front-end electronics'' Journal of Instrumentation 9 (02), (article id. C02015).







\end{thebibliography}







\end{document}